\documentclass{PoS}

\newcommand{\eq}[1]{\begin{equation}\label{#1}}
\newcommand{\en}{\end{equation}}
\newcommand{\ear}[1]{\begin{eqnarray}\label{#1}}
\newcommand{\enar}{\end{eqnarray}}
\newcommand{\tr}{{\rm{Tr}}}
\newcommand{\re}{{\rm{Re}}}
\newcommand{\dd}{{\rm{d}}}

\title{
\vspace*{-3cm}
\begin{flushright}\texttt{\footnotesize
CERN-PH-TH/2009-218}
\end{flushright}
\vfill
Dimensional reduction and the phase diagram of \\ 5d Yang-Mills theory}

\ShortTitle{Dimensional reduction and the phase diagram of 5d Yang-Mills theory}

\author{\speaker{A. Kurkela}\\
         Institute for Theoretical Physics,  ETH Z\"urich, CH-8093 Z\"urich, Switzerland\\
        E-mail: \email{kurkela@phys.ethz.ch}}

\author{Ph. de Forcrand\\
 Institute for Theoretical Physics,  ETH Z\"urich, CH-8093 Z\"urich, Switzerland\\
 and\\
 CERN, Physics Department, TH Unit, CH-1211 Geneva 23, Switzerland\\
        E-mail: \email{forcrand@phys.ethz.ch}}

\author{M. Panero\\
 Institute for Theoretical Physics,  ETH Z\"urich, CH-8093 Z\"urich, Switzerland\\
        E-mail: \email{panero@phys.ethz.ch}}

\abstract{We present a non-perturbative study of the phase diagram of 5d SU(2) Yang-Mills theory with one compact extra dimension on the lattice. Assuming at least a modest scale separation between the cutoff and the compactification scales leads to an \emph{exponential} separation between the compactification scale and the four-dimensional correlation length. While we demonstrate that it is not possible to take a full five-dimensional continuum limit, this dynamical generation of scale hierarchy opens up the possibility for us to make limited, but non-perturbative, predictions about continuum theories whose low-energy sector is described by an effective 5d Yang-Mills theory.
}

\FullConference{The XXVII International Symposium on Lattice Field Theory - LAT2009\\
		 July 26-31 2009\\
		 Peking University, Beijing, China}

\begin{document}

\section{Introduction}
Gauge theories defined in a spacetime with compact extra dimensions offer a very interesting platform for model building and phenomenology in physics beyond the Standard Model \cite{ArkaniHamed:2001ca}. However, at the same time the non-renormalizable nature of gauge theories in $d>4$ makes it notoriously difficult to extract any non-perturbative information from these theories. In fact, the extra-dimensional gauge theory is to be understood only as a low-energy effective description of a more fundamental theory, and therefore as a theory defined with a (momentum) cutoff $\Lambda$ indicating the scale where the effective description breaks down and the details of the underlying theory become significant. For a given compactification scale  $L_5$ of the extra dimension, this effective theory description makes sense only if there is at least a modest scale separation between the inverse cutoff $\Lambda^{-1}$ and the compactification length, $L_5 \Lambda \gg 1$. 

In these proceedings we report on our recent study~\cite{Kurkela}, in which we investigate the low-energy sector of the effective theory by trading the physical cutoff for a lattice regularization. In the cases where the details of the cutoff do not play a significant role, this allows us to obtain robust non-perturbative predictions from numerical simulations.

\section{Basic setup of the model}
As a prototype for an extra-dimensional gauge model, we study SU(2) Yang-Mills theory in 5d Euclidean spacetime, where one of the dimensions is taken to be periodic with a finite extent $L_5$ and the gauge fields are taken to obey periodic boundary conditions along the compact direction. 
The model is formally defined by the path integral
\begin{equation}
Z = \int\mathcal{D}A \textrm{e}^{-\mathcal{S}_E}, \quad \quad \mathcal{S}_E = \int {\dd}^4x  \int_0^{L_5}{\dd} x_5 \frac{1}{2 g_5^2 }\tr F_{MN}^2,\label{CS}
\end{equation}
with the indices $M,N$ running from 1 to 5. As this model is non-renormalizable, the continuum action in Eq.~(\ref{CS}) defines it only up to a regularization with an associated cutoff $\Lambda$. Note that in 5d the (bare) coupling $g_5^2$ has the dimension of a length, and the model can thus be parametrized by two dimensionless ratios, namely $L_5\Lambda$ and $g^2_5 \Lambda$. 

In order to study the model at non-zero coupling we regularize it on an anisotropic lattice with a lattice spacing $a$ in the four usual directions, letting the lattice spacing $a_5$ in the fifth direction take an independent value, such that $L_5=a_5N_5$. We take as our lattice action the Wilson action
\begin{eqnarray}
\hspace*{-0.4cm}
\mathcal{S}_E^L&=&\frac{\beta_5}{\gamma} \sum_{x,1 \le M < N \le 4} \left[1 - \frac{1}{2} \re\tr P_{MN}(x)\right]+\gamma \beta_5 \sum_{x,M=1}^{4} \left[ 1 - \frac{1}{2} \re\tr P_{M5}(x)\right], \; \mbox{with }
\beta_5=\frac{4 a}{g_5^2}.
\label{eq:lattice_action}
\end{eqnarray}
In the weak coupling limit, the anisotropy factor $\gamma$ reduces to the ratio of the lattice spacings
\begin{equation}
\lim_{\beta_5\rightarrow \infty} \frac{a}{a_5}=\gamma.
\end{equation}

An attempt to take the cutoffs to zero $(a,a_5) \rightarrow 0$ leads to (power-like) divergences, which cannot be absorbed in coefficients of an action with a finite number of terms, and thus the cutoff cannot be removed. Nevertheless, one can take the lattice spacing in \emph{one} of the directions to zero, $a_5\rightarrow0$, while keeping $N_5/\gamma\equiv \tilde{N}_5\approx L_5/a$ and the lattice spacing $a$ in the four usual directions fixed, since this corresponds to studying quantum mechanics of a 4d lattice with non-zero lattice spacing $a$. In the following we adopt this strategy for two reasons: \emph{(i)} Taking the continuum limit in the fifth direction reduces the set of dimensionless parameters characterizing the model from ($N_5,\gamma,\beta_5$) to ($\tilde{N}_5,\beta_5)$, such that the lattice model is parametrized by two dimensionless numbers in correspondence with the continuum theory. \emph{(ii)} At the same time, it reduces the cutoff effects in systems where the ratio $L_5/a$ is not very large. This turns out to be very important in practical simulations, since the 4d correlation length of the system strongly depends on this ratio.

After taking the continuum limit in the fifth dimension, one cannot proceed further to remove the remaining lattice spacing $a$.
However, the continuum limit can be approached in the sense that if the correlation length of the lattice model $\xi$ becomes very large compared to the lattice spacing $a$, the low-energy sector of the model becomes insensitive to the details of the discretization and the continuum symmetries get restored. 

\section{Phase diagram}
\begin{figure}
\begin{center}
\includegraphics*[width=7cm]{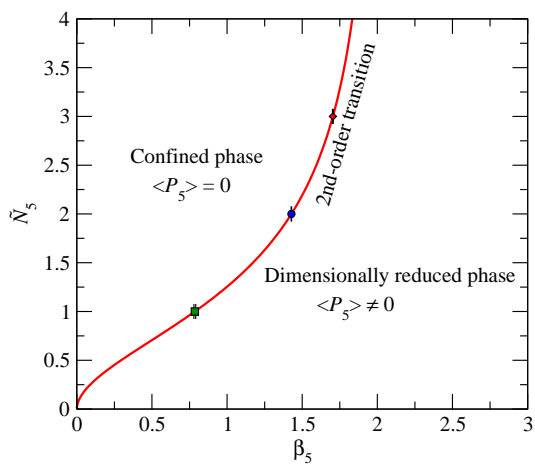}
\end{center}
\vspace*{-.7cm}
\caption{The phase diagram of the SU(2) gauge theory Eq.~(\protect\ref{eq:lattice_action}) in the $N_5\rightarrow \infty$ limit. The solid curve, drawn to guide the eye, goes through the origin with diverging derivative as expected from the strong coupling expansion and tends to an asymptotic value as $\tilde{N}_5\rightarrow \infty$. The confined phase is not analytically connected to the weak-coupling regime at $\beta_5\rightarrow \infty$ and is a pure artifact of the lattice regularization. }
\label{phasediag}
\end{figure}

As a first step to understand our model, we start by mapping out its phase diagram in the $(\beta_5,\tilde{N}_5)$-plane. For the isotropic uncompactified case the phase diagram, first investigated by Creutz~\cite{Creutz}, exhibits a confining phase at strong coupling and a Coulomb phase at weak coupling. We found this to be the case also for the system with anisotropic action Eq.~(\ref{eq:lattice_action}). The confining phase is not connected to the weak coupling regime and thus it is merely a lattice artifact.

In the Coulomb phase, the correlation length is infinite, implying that any compactification radius will lead to spontaneous breaking of the center symmetry along the fifth direction. We call this the dimensionally reduced phase for reasons which will become clear in the next Section.

Fig.~\ref{phasediag} shows the phase diagram of our model in the $(\beta_5,\tilde{N}_5)$-plane in the $N_5 \rightarrow \infty$ (i.e. $a_5 \rightarrow 0$) limit, obtained mostly from $4^4\times N_5$ lattices.
For any given value of $\tilde{N}_5$, one finds a confining phase and a dimensionally reduced phase, which are separated by a second-order phase transition in the trivial 4d Ising universality class expected for a (4+1)d, $\rm{Z}_2$-restoring transition where the correlation length of the Polyakov loop in the fifth direction diverges. We note that this is in contrast with the first-order transition observed by Creutz and is due to the weakening of the transition by the anisotropic lattice spacing in the fifth direction: increasing $N_5$ at fixed $\tilde{N}_5$ smoothens the transition out and eventually turns it into a second-order one.

\section{Dimensional reduction}
In this Section we study the dimensionally reduced phase, first in the weak (bare) coupling limit $\beta_5\rightarrow \infty$, then by numerical lattice simulations demonstrating that the description obtained in the weak coupling limit can be extended to non-zero couplings, up to corrections whose role will be discussed. In particular, we introduce an asymptotically free dimensionally reduced renormalizable effective theory for the long distance correlators, whose known properties can be directly translated to describe the properties of the 5d theory.

At separations $\Delta x$ large compared to the compactification scale, $\Delta x\gg L_5$, an   effective \linebreak SU(2) + adjoint Higgs theory for the static Kaluza-Klein modes of the gauge fields can be written 
\begin{equation}
 \mathcal{S}_{eff} = \frac{1}{g_4^2}\int {\dd}^4x  \left( \frac{1}{2}\tr F_{\mu\nu}^2 + \tr \left[ D_\mu A_5\right]^2 + m_5^2 \tr A_5^2+\lambda \tr A_5^4  \right) + \delta \mathcal{S} ,
\end{equation}
where the last term $\delta \mathcal{S}$ collectively denotes higher order terms in the gauge fields, whose effects are suppressed by powers of $m_5 L_5$ and turn out to be irrelevant for our purposes.
 
The parameters of the effective theory can be estimated in the weak coupling limit by requiring that, in the regime where the effective theory is expected to give a good description of the full theory, the correlators of the full and of the effective theory match to a given order in the bare coupling $g_5$. To leading order this gives for the renormalized 4d coupling constant and mass of $A_5$
\begin{eqnarray}
g_4^2(L_5)&=&\frac{g^2_5}{L_5}, \quad
m_5^2(L_5)\propto\frac{g_5^2}{L_5}\frac{1}{L_5^2}.
\end{eqnarray}
While at leading order the parameters of the effective theory are functions of $L_5$ and $g_5^2/L_5$ only, at higher loop orders the divergences in the 5d theory give dependence also on $g_5^2/a$.

In contrast with the gauge fields in the fifth direction, the static Kaluza-Klein modes of the gauge fields in the usual four directions do not perturbatively obtain mass. As a consequence, at length scales $\Delta x\gg m_5^{-1}$ the theory is described by a four-dimensional continuum pure gauge theory
\begin{equation}
\mathcal{S}_{eff} =\int {\dd}^4x   \frac{1}{2g_4^2}\tr F_{\mu\nu}^2.
\label{MQCD}\end{equation}
The four-dimensional Yang-Mills theory is well-known to non-perturbatively generate a gluon mass, implying a non-zero four dimensional string tension for the full five-dimensional theory
\begin{equation}
\sigma_{4d}  \sim \frac{1}{L_5^2} \exp \left[ - \frac{1}{b_0 g_4^2(L_5)}\right] \sim \frac{1}{a^2 \tilde{N}_5^2} \exp \left[ - \frac{\beta_5\tilde{N}_5}{4b_0} \right],  
\label{strng}
\end{equation}
with $b_0=11/24\pi^2$ for SU(2).

\begin{figure}
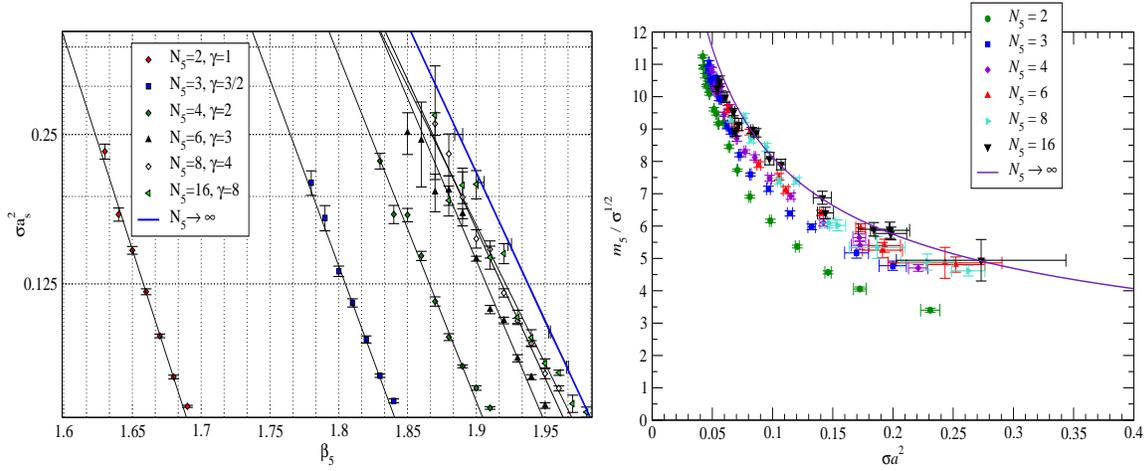

\begin{center}
\includegraphics*[width=7.8cm,height=5.83cm]{stringtension.eps}\hfill
\includegraphics*[width=7.2cm,height=6.2cm]{a5mass_vs_stringtension2_L2.eps}
\end{center}
\vspace{-0.5cm}
\caption{Left: String tension measured from a set of $6^4\times N_5$ lattices with fixed $\tilde{N}_5$. In the limit $N_5 \rightarrow \infty$, the data become consistent with Eq.~(\protect\ref{strng}). Right: Simulation results for the mass of the $A_5$ field in units of the square root of the 4d string tension, versus the string tension in units of the lattice spacing. Data are obtained at fixed $N_5/\gamma=2$ and for different $N_5$ values. The solid curve shows the extrapolation to the $N_5 \to \infty$ limit.}
\label{stringtension}
\end{figure}

We have numerically checked the behavior predicted by Eq.~(\ref{strng}). The results are shown in the left panel of Fig.~\ref{stringtension}, where the string tension extracted from torelon correlators is displayed for fixed $\tilde{N}_5=2$ and increasing $N_5$. While for small $N_5$, the fitted coefficient in the exponential is quite large, approaching the continuum limit in the fifth direction makes the data consistent with Eq.~(\ref{strng}).

In this setup dimensional reduction takes place, not only when $\tilde{N}_5\rightarrow 0$\footnote{We do not consider this case further as in this limit the extent of the extra dimension is smaller than the lattice spacing and cannot possibly describe the low-energy sector of a five-dimensional continuum theory.}, but also in a highly counter-intuitive way when the extent of the fifth direction becomes large in units of the cutoff, i.e., $\tilde{N}_5\approx L_5/a\gg1$. Consider Eq.~(\ref{strng}) keeping the four-dimensional string tension $\sigma_{4d}$ and the bare coupling $\beta_5$ of the 5d theory fixed. It implies that increasing $\tilde{N}_5$ linearly
results in an exponential increase in the 4d correlation length $\xi_{4d}/a\equiv 1/\sqrt{\sigma_{4d} a^2 } \sim \tilde{N}_5 \exp(+\beta_5 \tilde{N}_5/8b_0)$, 
rendering the extent of the fifth dimension in units of 4d measurable quantities
\begin{equation}
L_5/\xi_{4d}\sim \exp(- \beta_5 \tilde{N}_5/8b_0)
\end{equation}
 vanishingly small in the limit $\tilde{N}_5\rightarrow \infty$. That is, when the extent of the fifth dimension is taken to infinity in units of the ultraviolet cutoff, it goes to zero in units of a four-dimensional observer~\cite{Chandrasekharan:1996ih}.

\section{Approaching the continuum}
\begin{figure}
\begin{center}
\includegraphics*[width=9.2cm]{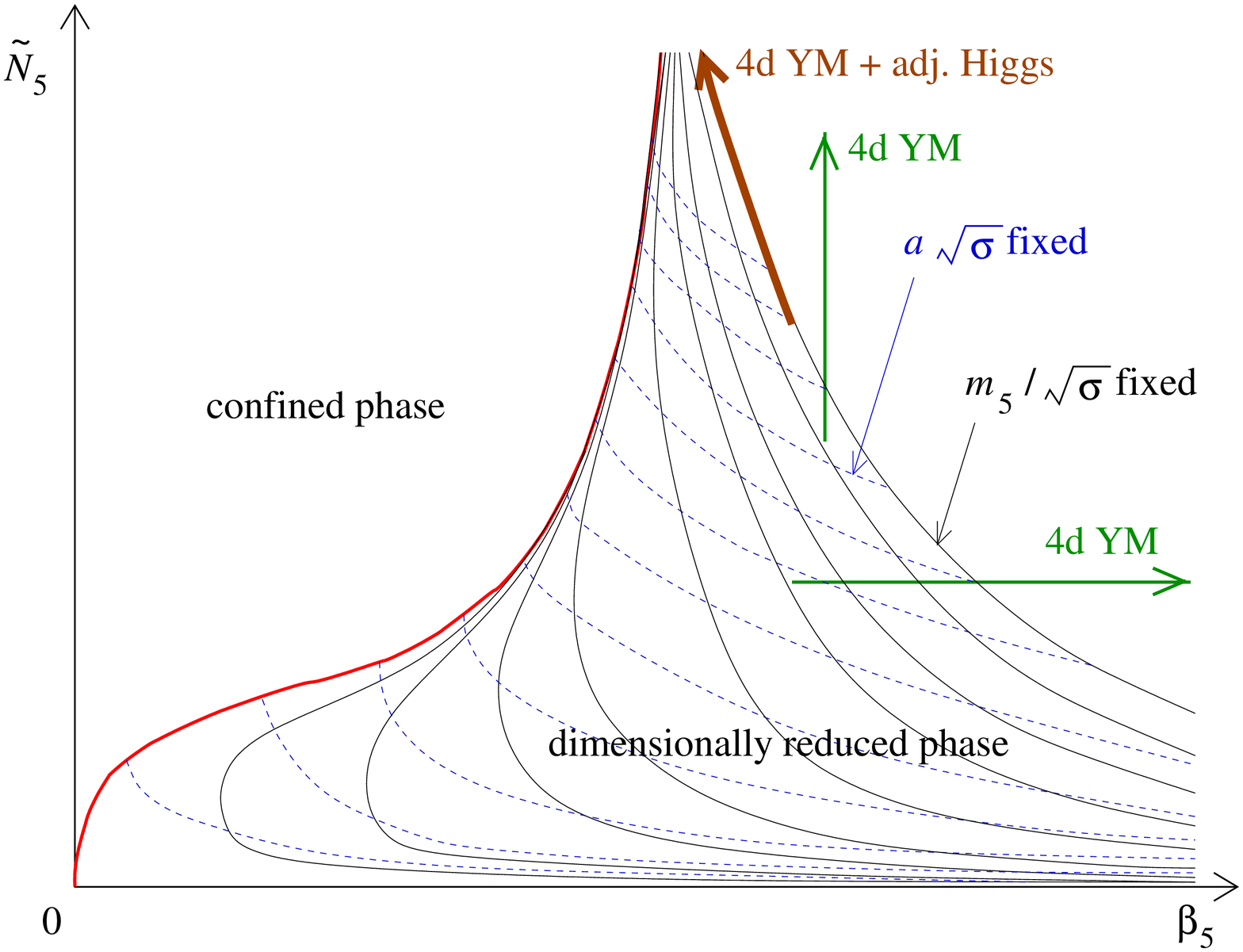}
\end{center}
\vspace{-.72cm}
\caption{Sketch of the lines of constant $\xi_{4d}m_5$ (solid black lines) and $\sigma_{4d} a^2$ (dashed blue lines). At large $\beta_5$, the lines of constant $\xi_{4d}m_5$ tend to hyperbolas $\beta_5 N_5/\gamma=\mbox{const}$. When moving up along each line of fixed $\xi_{4d}m_5$ (brown arrow), successive lines of constant $\sigma_{4d} a^2$ are crossed and the 4d lattice spacing is reduced. The 4d lattice spacing can be reduced also by increasing $\beta_5$ or $\tilde{N}_5=N_5/\gamma$. In these cases, lines of constant $\xi_{4d}m_5$ are crossed, implying the decoupling of the gauge field $A_5$. }
\label{LCP_sketch_with_a}
\end{figure}

In this Section, we discuss how the results of the previous Sections can be used to approach the continuum in the sense of letting the correlation lengths become large compared to the lattice spacing in the 5d theory. First, we note 
that as $\tilde{N}_5$ and $\beta_5$ appear in symmetrical form in the effective theory of Eq.~(\ref{MQCD}) we can take a continuum limit either at fixed $\tilde{N}_5$ by taking $\beta_5\rightarrow \infty$, or as well by keeping the bare coupling constant $\beta_5$ fixed and taking $\tilde{N}_5\rightarrow \infty$, marked by the two green arrows in Fig.~\ref{LCP_sketch_with_a}. This does not contradict the non-renormalizability of the 5d theory, since in this limit the five-dimensional degrees of freedom decouple and the target continuum theory will be nothing but the usual renormalizable 4d pure gauge theory. 

On the other hand, in order to investigate the more interesting case of a theory which still includes features derived from the original five-dimensional model while letting the correlation lengths diverge in lattice units, one can use the freedom of tuning the parameters of the bare theory. 
This can be accomplished by letting $\xi_{4d}/a\rightarrow \infty$ while keeping $\xi_{4d}m_5$ fixed. In the weak coupling limit this is achieved by keeping $g_5^2/L_5$ (or equivalently $\tilde{N}_5 \beta_5$) fixed and taking $\tilde{N}_5\rightarrow\infty$. These lines would correspond to hyperbolas in Fig.~\ref{LCP_sketch_with_a}. However, with non-zero coupling the matching coefficients acquire also dependence on $g_5^2/a$ (or on $\beta_5$), deforming 
the hyperbolas. These lines can be tracked either by computing the matched coefficients to higher orders in perturbation theory, or even better by determining the mass non-perturbatively, as depicted in the right panel of Fig.~\ref{stringtension}.

Due to the second-order nature of the phase transition between the dimensionally reduced and confined phases, the lines of fixed $\xi_{4d}m_5$ cannot intersect the phase transition line where $m_5\rightarrow 0$. As a consequence, close to the transition the lines of fixed $\xi_{4d} m_5$ (solid lines in Fig.~\ref{LCP_sketch_with_a}) will bend along the transition line. This allows tuning the correlation length of the static Kaluza-Klein mode to arbitrarily large values for any given $\xi_{4d}m_5$. However, at the phase transition it is only the static KK-mode of the gauge field in the fifth direction that diverges, eventually decoupling the rest of the KK-tower in this continuum limit. Again, the continuum target theory is four-dimensional, namely SU(2)\,+\,adjoint Higgs (arrow in Fig.~3), and there is no contradiction with non-renormalizability.

\section{Conclusions and implications for phenomenology}
We have presented a study of 5d SU(2) gauge theory on a lattice with a compact dimension of size $L_5$. The theory is non-renormalizable, but 
the dynamical generation of a scale hierarchy between the cutoff scale, the compactification scale, and the four-dimensional correlation length allows one to make non-perturbative predictions about the low-energy sector of the continuum theory using numerical lattice simulations.
We used anisotropic lattices to accommodate the large correlation lengths in the four usual directions and to reduce the discretization errors in the fifth direction. This turned out to be essential for checking numerically the scaling of the four-dimensional string tension as given by Eq.~(\ref{strng}). While the simulations we performed are in agreement with those of Ref.~\cite{Ejiri:2000fc}, our conclusions differ significantly. In particular, our results  demonstrate that all attempts to take the cutoff scale to infinity inevitably lead to continuum target theories which are four-dimensional. A five-dimensional continuum limit, where the correlation lengths of an arbitrary number of KK-modes would diverge in units of the cutoff scale, cannot be reached.

Our strategy can be employed to produce phenomenologically interesting non-perturbative predictions of continuum theories. In particular, Eq.~(\ref{strng}) yields a robust upper bound for the radius of the extra dimension, with the only assumption that the underlying theory can be described by a 5d gauge theory at the compactification scale\footnote{This can be seen in Fig.~\ref{LCP_sketch_with_a}: along a dashed line corresponding to a given value of the cutoff, i.e. $\sigma a^2$, the ratio $\tilde{N}_5 \approx L_5/a$ cannot exceed the maximum value reached at the intersection with the phase transition line.}. For example, taking the electroweak scale as the 4d scale and assuming the cutoff to be at the Planck scale gives a maximum compactification length for the extra dimension of the order of 10$M_P^{-1}$.
 Another application is the possible detection of the static KK-mode of the gauge field in the fifth direction. In this case, our strategy can be used to extract a non-perturbative estimate for the onset scale of new physics $\Lambda$. If the mass $m_5$ and the self-coupling $\lambda$ of the scalar Higgs-like particle can be determined, say, at future LHC experiments, this information can be non-perturbatively mapped via lattice simulations to the two parameters of our lattice model $\beta_5,\tilde{N}_5$, from which $L_5$ and $\Lambda$ can be uniquely determined.

In the present study we have only considered the simplest possible non-Abelian 5d model. While many of our predictions generalize to more realistic models, possible future extensions of this work include its extensions to higher rank groups, accommodating e.g., the Hosotani mechanism, and a systematic classification of the effects of the boundary conditions. Finally, our five-dimensional model also provides a natural setup for the inclusion of light fermions in a domain-wall-like construction. 

{\bf Acknowledgements:} PdF thanks KITPC for hospitality. 
AK acknowledges support from the SNF grant 20-122117 and MP from INFN. The simulations were performed at the Center for Scientific Computing (CSC), Finland under HPC-Europa2 Transnational Access programme.
We thank K.N.~Anagnostopoulos, K.~Farakos, L.~Giusti, M.~Golterman, K.~Kajantie, T.~Kennedy, M.~Laine, S.~Nicolis, K.~Rummukainen, C.~Scrucca, A.~Tsapalis and U.-J.~Wiese for discussions.

\end{document}